\documentclass{article}

\usepackage{graphicx,hyperref}

\usepackage{amsthm,amsfonts,amsopn,amsmath,amssymb,natbib}
\usepackage[latin1]{inputenc}
\usepackage{xspace}
\usepackage{float}
\usepackage{algorithm,algorithmic,color}
\usepackage[section]{placeins}
\usepackage{sidecap}
\usepackage[top=1.93in, bottom=1.6in, left=1.6in, right=1.6in]{geometry}
\usepackage{microtype}

\DeclareMathOperator{\Var}{Var}

\DeclareMathOperator{\D}{D}
\DeclareMathOperator{\Cov}{Cov}

\newcommand{\npar}{\par \vspace{2.3ex plus 0.3ex minus 0.3ex}}
\newcommand{\E}{\mathbb{E}}

\theoremstyle{plain}

\def\cov{\mathop{\rm Cov}} 
\def\E{\mathbb{E}} 

\author{Tim Salimans\footnote{\href{mailto:salimanstim@gmail.com}{salimanstim@gmail.com}} \and David A. Knowles\footnote{Stanford University}}
 
\title{On Using Control Variates with Stochastic Approximation for Variational Bayes and its Connection to Stochastic Linear Regression}

\begin{document}

\renewcommand{\thefootnote}{\fnsymbol{footnote}}

\maketitle

\renewcommand{\thefootnote}{\arabic{footnote}}

\begin{abstract} Recently, we and several other authors have written about the possibilities of using stochastic approximation techniques for fitting variational approximations to intractable Bayesian posterior distributions. Naive implementations of stochastic approximation suffer from high variance in this setting. Several authors have therefore suggested using control variates to reduce this variance, while we have taken a different but analogous approach to reducing the variance which we call stochastic linear regression. In this note we take the former perspective and derive the ideal set of control variates for stochastic approximation variational Bayes under a certain set of assumptions. We then show that using these control variates is closely related to using the stochastic linear regression approximation technique we proposed earlier. A simple example shows that our method for constructing control variates leads to stochastic estimators with much lower variance compared to other approaches.
\end{abstract}

\section{Stochastic Approximation for Fixed-Form Variational Bayes}
The goal of fixed-form Variational Bayesian posterior approximation is to approximate an intractable posterior distribution $p(x|y)$ with a parameterized approximating distribution $q_{\eta}(x)$ of a given, more convenient, form. Here $\eta$ denotes the parameters of the approximation $q$, and $x$ denotes the unknowns (parameters, latent variables) of our model for which we would like to approximate the posterior distribution. The approximation is then determined by minimizing the KL-divergence between $q$ and $p$:

\begin{equation}
\label{eq:post2}
\hat{q} = \arg\min_{q(x)} \D[q|p] = \arg\min_{q(x)} \E_{q(x)}\left[\log \frac{q(x)}{p(x,y)}\right], 
\end{equation}

where $p(x,y)$ denotes the unnormalized posterior distribution $p(x|y)p(y)$. 
\npar
Numerically solving \eqref{eq:post2} requires us to somehow evaluate the expectation with respect to $q$. We can only do so analytically for a very limited set of approximations and posteriors, so several authors \citep{paisleyvb, blackbox, nottstochastic, durk, hoffman2012stochastic, wingateweber} have recently proposed addressing this expectation using Monte Carlo techniques. Specifically, \citet{paisleyvb} and others find that

\begin{equation}
\label{eq:stochgrad}
\nabla_{\eta}[\D(q|p)] = \E_{q_{\eta}}\left\{ \nabla_{\eta}[\log q_{\eta}(x)](\log q_{\eta}(x) - \log p(x,y)) \right\},
\end{equation}

where $\nabla_{\eta}[\D(q|p)]$ denotes the column vector gradient of the KL-divergence. They then propose to evaluate the expectation in this expression using Monte Carlo. Doing so gives unbiased stochastic estimates of the gradient of the KL-divergence which can be used in a stochastic optimization procedure as pioneered by \citet{robbinsmonro}. However, a naive Monte Carlo approximation of \eqref{eq:stochgrad} typically has too much variance to be of practical use. Most of the aforementioned authors therefore propose the use of variance reduction techniques, and specifically control variates, to make this approach work. We take another approach, based on the idea of `noise cancellation' in linear regression.

\section{Variance reduction by linear regression}
To approximate \eqref{eq:stochgrad} we first rewrite it as follows:

\begin{eqnarray}
\nabla_{\eta}[\D(q|p)] & = & \E_{q_{\eta}}\left\{ \nabla_{\eta}[\log q_{\eta}(x)](\log q_{\eta}(x) - \log p(x,y)) \right\} \\
& = & \E_{q_{\eta}}\left\{ (\nabla_{\eta}[\log q_{\eta}(x)] - \E_{q} \nabla_{\eta}[\log q_{\eta}(x)])(\log q_{\eta}(x) - \log p(x,y)) \right\} \\
& = & \Cov_{q}[\nabla_{\eta} \log q_{\eta}(x), \log q_{\eta}(x) - \log p(x,y)]\label{eq:cov},
\end{eqnarray}

where we make use of the fact that $\E_{q} \nabla_{\eta}[\log q_{\eta}(x)] = 0$ for any $q$ \citep[see e.g.][]{blackbox}. This covariance \eqref{eq:cov} can be approximated without bias by its sample estimator using samples drawn from $q$:

\begin{eqnarray}
x^{*}_{1},x^{*}_{2},\ldots,x^{*}_{S} & \sim & q_{\eta}(x) \\
\hat{g}_{\text{cov}} & = & \hat{\Cov}_{q}[\nabla_{\eta} \log q_{\eta}(x), \log q_{\eta}(x) - \log p(x,y)] \\
& = & \frac{1}{S-1} \sum_{i=1}^{S} [\nabla_{\eta} \log q_{\eta}(x^{*}_{i}) - \hat{m}][\log q_{\eta}(x^{*}_{i}) - \log p(x^{*}_{i},y)] \label{eq:covest}\\
\hat{m} & = & \frac{1}{S} \sum_{i=1}^{S} \nabla_{\eta} \log q_{\eta}(x^{*}_{i}).
\end{eqnarray}

The estimator \eqref{eq:covest} will generally already have much lower variance than when approximating \eqref{eq:stochgrad} directly. Nevertheless, we can reduce the variance further by approximating the gradient using

\begin{eqnarray}
\hat{g}_{\text{reg}} & = & \Cov_{q}[\nabla_{\eta} \log q_{\eta}(x),\nabla_{\eta} \log q_{\eta}(x)]\hat{g}_{\text{nat}}, \text{ with } \label{eq:natgrad}\\
\hat{g}_{\text{nat}} & = & \hat{\Cov}_{q}[\nabla_{\eta} \log q_{\eta}(x),\nabla_{\eta} \log q_{\eta}(x)]^{-1}\hat{g}_{\text{cov}} \label{eq:ngalpha} \\
& = & \hat{\Cov}_{q}[\nabla_{\eta} \log q_{\eta}(x),\nabla_{\eta} \log q_{\eta}(x)]^{-1}\hat{\Cov}_{q}[\nabla_{\eta} \log q_{\eta}(x), \log q_{\eta}(x) - \log p(x,y)], \nonumber
\end{eqnarray}

where $\hat{g}_{\text{nat}}$ is a stochastic estimate of the \textit{natural gradient}, and where the two covariances terms in the last line are estimated using the same random draws $x^{*}_{1},x^{*}_{2},\ldots,x^{*}_{S}$. In \eqref{eq:natgrad}, we multiply the natural gradient with the \textit{exact} covariance, rather than its estimate. Analytic expressions for this covariance are indeed available for many common choices of tractable exponential family $q(x)$. When an analytical expression is not available, this covariance can be calculated efficiently using quadrature as long as the partial derivatives $\frac{\partial}{\partial x_{i}} \log q_{\eta}(x)$ only depend on a low dimensional subset of $x$, for example when $q(x)=\prod_{j}q(x_{j})$ is a factorized distribution. Note that the availability of this covariance does not in any way depend on the form of the posterior $p(x,y)$ that is being approximated, and is therefore `black box' in the terminology of \citet{blackbox}.
\npar
The estimator \eqref{eq:natgrad} has much lower variance than our original one \eqref{eq:covest} because the noise in our estimates of the two covariance terms largely cancels each other out. This is the same effect that causes the estimator $(X'X)^{-1}X'y$ to be efficient for classical linear least squares regression, in contrast to theoretical alternatives like $\E[X'X]^{-1}X'y$ that are inefficient. 
\npar
Unlike the other estimators, $\hat{g}_{\text{reg}}$ is biased, and should therefore not be used in a stochastic gradient descent procedure directly. In our earlier work \citep{linregvb} we therefore propose an adaptation of stochastic gradient descent that eliminates this bias, but keeps the much lower variance of this estimator. An added advantage of this approach is that it does not require the analytic calculation of $\Cov_{q}[\nabla_{\eta} \log q_{\eta}(x),\nabla_{\eta} \log q_{\eta}(x)]$. Another strategy would be to try and see whether we can use the ideas behind this estimator to construct an unbiased estimator of the gradient. This is what we do in the Section~\ref{sec:newcv}, but before we do so we first present another method for approximating the covariance terms of \eqref{eq:ngalpha}.

\section{Differentiating the Monte Carlo sampler}
\label{sec:difmc}

First note that our derivation in \eqref{eq:cov} can be stated more generally as
\begin{equation}
\label{eq:cov_to_est}
\nabla_{\eta} \E_{q}[h(x)] = \Cov_{q}[\nabla_{\eta} \log q_{\eta}(x), h(x)],
\end{equation}
for any distribution $q_{\eta}(x)$ and function $h(x)$ such that this covariance exists. Now assume that we can approximate expectation $\E_{q}[h(x)]$ unbiasedly using draws $x_{i}^{*} = s(\eta,s_{i}^{*})$ from a pseudo random number generator $s()$ using random number seeds $s_{i}^{*}$:
\begin{equation}
\label{eq:montecarlo}
\hat{\E}_{q}[h(x)] = \frac{1}{S} \sum_{i=1}^{S} h[x_{i}^{*}] = \frac{1}{S} \sum_{i=1}^{S} h[s(\eta,s_{i}^{*})].
\end{equation}
If we then have that $x$ is a set of continuous variables and that $h(x)$ and $s(\eta,s_{i}^{*})$ are continuously differentiable, we also have that
\begin{equation}
\label{eq:gradcov} 
\hat{\Cov}_{q}[\nabla_{\eta} \log q_{\eta}(x), h(x)] = \nabla_{\eta} \frac{1}{S} \sum_{i=1}^{S} h[s(\eta,s_{i}^{*})]
\end{equation}
is an unbiased estimator for our covariance term \eqref{eq:cov_to_est}. This way, we can approximate the covariance terms of \eqref{eq:ngalpha} by differentiating through the Monte Carlo estimators of $\E_{q}[\nabla_{\eta} \log q_{\eta}(x)]$ and $\E_{q}[\log q(x) - \log p(x,y)]$. We find that this type of Monte Carlo estimator often has lower variance than when we simply use the sample covariance. In \citet{linregvb} we use this strategy successfully to approximate the natural gradient of the KL-divergence \eqref{eq:ngalpha} for approximations $q(x)$ in the exponential family. \citet{durk} have since derived the same principle independently, but then for the regular gradient \eqref{eq:stochgrad}, and used it to perform variational inference with an auto-encoding neural network.

\section{New Control Variates for Stochastic Approximation Variational Bayes}
\label{sec:newcv}
If our goal is to stochastically approximate an expectation $\E_{q} f(x)$ using Monte Carlo, we may equivalently approximate $\E_{q} f(x) - \alpha h(x)$ for any scalar $\alpha$ and any function $h(x)$ for which we know that $\E_{q} h(x) = 0$. If $f(x)$ and $h(x)$ are then positively correlated when sampling from $q$, and $\alpha$ is chosen appropriately, the resulting estimator will have lower variance than when approximating the original expression directly. This variance reduction technique is called the control variates method. We now propose a new set of control variates that may be used to decrease the variance of our unbiased stochastic estimator \eqref{eq:covest}.
\npar
For approximating the $i$-th component of the $k \times 1$ vector $\nabla_{\eta} \D(q|p)$, we propose to use the $1 \times k$ vector of control variates $h^{i}$, with

\begin{equation}
h^{i} = \hat{\Cov}_{q}\left[\frac{\partial}{\partial x_{i}} \log q_{\eta}(x), \nabla_{x} \log q_{\eta}(x)\right] - \Cov_{q}\left[\frac{\partial}{\partial x_{i}} \log q_{\eta}(x), \nabla_{x} \log q_{\eta}(x)\right], \label{eq:newcv}
\end{equation}

where the second term $\Cov_{q}$ denotes the exact (analytical) expression for the covariance, and the first term $\hat{\Cov}_{q}$ denotes its sample estimator. Using these control variates, we have

\begin{eqnarray}
\frac{\partial}{\partial \eta_{i}} \D(q|p) & = & \E_{q_{\eta}}[f^{i} - h^{i}\alpha^{i}], \text{ with} \label{eq:cg}\\
f^{i} & =& \hat{\Cov}_{q}\left[\frac{\partial}{\partial \eta_{i}} \log q_{\eta}(x), \log q_{\eta}(x) - \log p(x,y)\right],
\end{eqnarray}

where $\alpha^{i}$ is the $k \times 1$ vector of control variate coefficients. It remains for us to determine how the set this vector of coefficients. A standard result tells us that the ideal coefficients are given by

\begin{equation}
\alpha^{i*} = \Var_{q}[h^{i}]^{-1}\Cov_{q}[h^{i},f^{i}(x)],
\end{equation}

which in practice is approximated by

\begin{equation}
\hat{\alpha}^{i} = \hat{\Var}_{q}[h^{i}]^{-1}\hat{\Cov}_{q}[h^{i},f^{i}(x)], \label{eq:cv1}
\end{equation}

with the hat symbol again denoting sample estimators.

\section{Exponential family approximation and posterior}
\label{sec:expfam}
Up until now, all of our derivations have been for general distributions $q(x)$. In this section, we consider the special case in which we use an approximation $q(x)$ that is a member of the exponential family. That is, we assume

\begin{equation}
\label{eq:expfam}
\log q_{\eta}(x) = T(x)\eta - Z(\eta),
\end{equation}

with $T(x)$ a $1 \times k$ vector of sufficient statistics, and $Z(\eta)$ a normalizing constant. In this case we have

\begin{eqnarray}
\nabla_{x} \log q_{\eta}(x) & = & T(x) - \E_{q}[T(x)], \text{ and} \\
\Cov_{q}[\nabla_{\eta} \log q_{\eta}(x), \nabla_{\eta} \log q_{\eta}(x)] & = & \Cov_{q}[T(x), T(x)].
\end{eqnarray}

Using these expressions, we can rewrite \eqref{eq:cov} as

\begin{eqnarray}
\nabla_{\eta}[\D(q|p)] & = & \Cov_{q}[\nabla_{\eta} \log q_{\eta}(x), \log q_{\eta}(x) - \log p(x,y)] \\
& = & \Cov_{q}[T(x), \log q_{\eta}(x) - \log p(x,y)] \\
& = & \Cov_{q}[T(x), T(x)]\eta - \Cov_{q}[T(x),\log p(x,y)],
\end{eqnarray}

a result we also derived in our earlier work \citep{linregvb}.
\npar
Let us now also assume that the unnormalized posterior $p(x,y)$ is of the same exponential family form. That is,

\begin{equation}
\label{eq:expfam}
\log p(x,y) = T(x)\tilde{\eta}(y) + c(y),
\end{equation}

where $\tilde{\eta}$ is another set of (unknown) parameters and $c$ is constant in $x$. Both quantities would normally depend on the data $y$.
\npar
In this case, we have the following result for the ideal control variate coefficients $\alpha^{i*}$:

\begin{eqnarray}
\label{eq:linregalpha}
\alpha^{i*} & = & \Cov_{q}[\nabla_{\eta} \log q_{\eta}(x), \nabla_{\eta} \log q_{\eta}(x)]^{-1}\Cov_{q}[\nabla_{\eta} \log q_{\eta}(x), \log q_{\eta}(x) - \log p(x,y)] \nonumber\\
& = & \Cov_{q}[T(x), T(x)]^{-1}\Cov_{q}[T(x), \log q_{\eta}(x) - \log p(x,y)] \label{eq:regcv} \\
& = & \Cov_{q}[T(x), T(x)]^{-1}\Cov_{q}[T(x), T(x)\eta - T(x)\tilde{\eta}] = \eta - \tilde{\eta} \hspace{0.2cm} \forall i. \label{eq:idealc}
\end{eqnarray}

Hence all control variate coefficient vectors $\alpha^{i*}$ have the same ideal values, which are given by the `stochastic linear regression' of \eqref{eq:ngalpha}. In practice, the coefficients are once again estimated by plugging in sample estimators, but in this special case that does not change their value:

\begin{equation}
\hat{\alpha} = \hat{g}_{\text{nat}} = \hat{\Cov}_{q}[T(x), T(x)]^{-1}\hat{\Cov}_{q}[T(x), T(x)](\eta - \tilde{\eta}) = \eta - \tilde{\eta} = \alpha^{*}
\end{equation}

Substituting in the estimated control variate coefficients, our stochastic estimate of the gradient of the KL-divergence \eqref{eq:cg} becomes:

\begin{eqnarray}
\hat{\nabla}_{\eta_{i}} \D(q|p) & = & \hat{\E}_{q_{\eta}}[f - h\hat{\alpha}] \\
& = & \hat{\Cov}_{q}[T(x), \log q_{\eta}(x) - \log p(x,y)] \nonumber\\
&& - (\hat{\Cov}_{q}[T(x),T(x)] - \Cov_{q}[T(x),T(x)])\hat{\alpha} \label{eq:stochgradest} \\
& = & \hat{\Cov}_{q}[T(x), T(x)\eta - T(x)\tilde{\eta}] \nonumber\\
&& - (\hat{\Cov}_{q}[T(x),T(x)] - \Cov_{q}[T(x),T(x)])\hat{\alpha} \\
& = & \Cov_{q}[T(x),T(x)]\hat{\alpha} \\
& = & \Cov_{q}[T(x),T(x)]\hat{g}_{\text{nat}} \label{eq:exact} \\
& = & \Cov_{q}[T(x),T(x)](\eta-\tilde{\eta}).
\end{eqnarray}

Hence, our `stochastic' estimate of the gradient has in fact become a deterministic estimate with zero variance. This means that the chosen control variates are indeed optimal if $q$ and $p$ are of the same exponential family form. Note that \eqref{eq:exact} is identical to our `biased' estimator \eqref{eq:natgrad}, which in this special case is thus also exact.
\npar
The situation where $q$ and $p$ are of the same functional form is completely hypothetical and this will not be the case in practice. Nevertheless, we find that, even when $q$ and $p$ have completely different forms, the control variate coefficients in \eqref{eq:regcv} are generally still very close to optimal and the resulting stochastic gradient estimate \eqref{eq:stochgradest} still has much lower variance compared to other stochastic estimators. We illustrate this with a simple example in the next section.

\section{Toy example: logistic regression}
As an example, we consider approximating the gradient of

\begin{equation}
\nabla_{\eta} \E_{q_{\eta}}[\log q_{\eta}(x) - \log p(x,y)], \label{eq:to_approximate}
\end{equation}

with univariate $x$, approximation $q_{\eta}(x) = N(x; \mu,\sigma^{2})$, and unnormalized posterior

\begin{equation}
\log p(x,y) = \log p(x) = x - \log[1+\exp(x)]. \label{eq:logitprob}
\end{equation}

Terms of this form \eqref{eq:logitprob} occur in the likelihood of logistic regression models such as the one considered by \citet{paisleyvb}. We deliberately pick an improper posterior consisting of only a single likelihood term to make sure the functional forms of $q(x)$ and $p(x)$ are as different as possible. Note that this biases the experiment against our method, which was derived under the assumption that $q(x)$ and $p(x)$ are of approximately the same form.
\npar
We approximate \eqref{eq:to_approximate} using the `simple' stochastic approximation defined in \eqref{eq:stochgrad}, the covariance approximation in \eqref{eq:cov}, the covariance approximation using our control variates and the coefficients in \eqref{eq:cv1}, the covariance approximation using our control variates with the coefficients in \eqref{eq:regcv}, and the gradient based approximation in \eqref{eq:gradcov} using the same control variates. We also compare against the generic control variate strategy recently proposed by \citet{blackbox}, the delta method of \citet{paisleyvb} which calculates $\E_{q} \log q(x)$ analytically and which uses a 2nd order Taylor approximation as the control variate for $\log p(x)$, and the estimator of \citet{durk} discussed in Section~\ref{sec:difmc}. Finally, we also report the result of using our biased estimator $\hat{g}_{\text{reg}}$, using both the sample covariance estimates and the gradient based estimates from Section~\ref{sec:difmc}.
\npar
We evaluate each of the approximations using 50 samples from the approximate posterior. For the methods using control variates we subdivide this into 25 samples for setting the control variate coefficients and 25 for evaluating the gradient. Note that we need to use different random draws for both steps to ensure that our estimate of the gradient is unbiased. The combined procedure of optimizing the coefficients and estimating the gradient is repeated 100,000 times for each method. We report the mean squared error of the different estimators in Table~\ref{vartable}. The MATLAB code for this experiment is available at \url{gist.github.com/TimSalimans/8279968}.

\begin{table}[H] 
\caption{Mean squared errors of the different stochastic gradient estimators for different settings of the parameters of the approximate posterior ($\mu,\sigma^{2}$)} 
\centering 
\begin{tabular}{c c c c c c} 
\hline\hline 
stochastic approximation method & $\mu=0,\sigma^{2}=2$ & $\mu=-2,\sigma^{2}=2$ & $\mu=2,\sigma^{2}=2$  & $\mu=0,\sigma^{2}=4$ \\ [0.5ex] 
\hline 
simple approximation \eqref{eq:stochgrad} & 0.5194 & 0.4242 & 2.2606 & 1.9734 \\
covariance approximation \eqref{eq:covest} & 0.3238 & 0.3524 & 0.8273 & 1.3296\\
\textbf{covariance + `ideal' c.v.} \eqref{eq:cv1} & \textbf{0.0060} & \textbf{0.0172} & \textbf{0.0179} & \textbf{0.0978}\\
\textbf{covar. + `regression' c.v.} \eqref{eq:regcv} & \textbf{0.0066} & \textbf{0.0233} & \textbf{0.0234} & \textbf{0.1147}\\
\textbf{covar + `ideal' c.v. + grad \eqref{eq:gradcov}} & \textbf{0.0010} & \textbf{0.0023} & \textbf{0.0023} & \textbf{0.0136}\\
\citet{blackbox} c.v. & 0.6133 & 0.6764 & 1.2663 & 3.0090\\
delta method \citet{paisleyvb} & 0.0252  &  0.0111  &  0.0240  &  0.4968 \\
\citet{durk}, \eqref{eq:gradcov} & 0.0472  &  0.0888  &  0.1930  &  0.1499 \\
\hline
\textbf{$\hat{g}_{\text{reg}}$ + sample cov. \eqref{eq:natgrad} \eqref{eq:covest} } & \textbf{0.0009}  &  \textbf{0.0062}  &  \textbf{0.0062}  &  \textbf{0.0180} \\
\textbf{$\hat{g}_{\text{reg}}$ + gradient \eqref{eq:natgrad} \eqref{eq:gradcov} } & \textbf{0.0006}  &  \textbf{0.0032}  &  \textbf{0.0032}  &  \textbf{0.0101} \\
\hline 
\end{tabular} 
\label{vartable} 
\end{table}

Table \ref{vartable} shows that, among the unbiased stochastic estimators, the control variates proposed in Section~\ref{sec:newcv} give the most accurate estimates for most settings of $\mu,\sigma^{2}$. The control variate strategy of \citet{blackbox} does improve upon the simple estimator \eqref{eq:stochgrad} for the same number of samples, but it loses this advantage when reserving half of the samples for calculating the control variate coefficients. An advantage of the control variate strategy of \citet{blackbox} is that it does not assume we can calculate $\Cov_{q}\left[\frac{\partial}{\partial x_{i}} \log q_{\eta}(x), \nabla_{x} \log q_{\eta}(x)\right]$. Our proposed algorithm in \citet{linregvb} also does not require this. The delta method control variate proposed in \citet{paisleyvb} and the estimator of \citet{durk} both outperform the simple estimators by making use of the structure (gradient/hessian) of $\log p(x)$. Nevertheless, both are outperformed in this setting by the simple versions of our control variate strategy which are agnostic to the form of $\log p(x)$. When we also use the gradient of $\log p(x)$ via \eqref{eq:gradcov} we are able to reduce the variance further by an order of magnitude. 
\npar
In terms of accurately estimating the gradient, our regression based estimate $\hat{g}_{\text{reg}}$ using the sample covariance is clearly the best among the `black box' estimators that do not rely on derivations relating to $\log p(x)$. Here we find that the squared bias contributes about 20\% on average to the mean squared error for this estimator. The squared bias decreases quadratically in the number of samples \citep{linregvb}, so this fraction will be lower if more samples are used. The downside of using a biased estimator in a stochastic gradient descent procedure is that the error in the gradient estimate does not average out over multiple iterations. This can be solved by using an adaptation of stochastic gradient descent \citep{linregvb}.
\npar
The gradient based estimator for $\hat{g}_{\text{reg}}$ based on Section~\ref{sec:difmc} performs about equally well compared to our (unbiased) control variate strategy when using the gradient. Interestingly, we find that $\hat{g}_{\text{reg}}$ is also nearly unbiased, so the two estimators are comparable in this regard as well.

\section{Conclusion}
In this note we have introduced new control variates for stochastic approximation of the gradient of the KL-divergence between an approximating distribution $q(x)$ and an intractable target $p(x|y)$. We have shown empirically that these control variates greatly reduce the variance in our stochastic estimates. Furthermore, we have shown a connection between these control variates and our `stochastic linear regression' framework proposed in \citet{linregvb}. In that work we extend our results to approximations for which we cannot calculate $\Cov_{q}[\nabla_{\eta} \log q_{\eta}(x), \nabla_{\eta} \log q_{\eta}(x)]$ analytically or using quadrature. In addition, we propose using the \textit{natural gradient} for minimizing the KL-divergence, which we find to provide a much better search direction than the regular gradient.
\npar
An apparent disadvantage of our approach is that it requires the inversion of our estimate of the $\cov_{q}[\nabla_{\eta} \log q_{\eta}(x), \nabla_{\eta} \log q_{\eta}(x)]$ matrix. However, note that this covariance matrix is sparse if elements of $x$ are independent under our posterior approximation, and that we can therefore also approximate it with a sparse matrix. Specifically, if we have a fully factorized posterior approximation $q(x)=\prod_{i}q(x_{i})$ with two parameters per $q(x_{i})$ as in e.g. \citet{blackbox}, our method only requires the inversion of matrices of size $2 \times 2$. Finally, note that the derivations in this paper (excluding Section~\ref{sec:expfam}) apply to general approximation $q(x)$ and not just those in the exponential family. Taking the contributions here together with our earlier work \citep{linregvb}, our approach is now general enough to be applied to all of the examples in all of the papers we have cited in this note.

\clearpage
\bibliographystyle{ba}
\bibliography{biball}

\end{document}